\documentclass[conference]{IEEEtran}
\IEEEoverridecommandlockouts
\usepackage{cite}
\usepackage{amsmath,amssymb,amsfonts}
\usepackage{algorithmic}
\usepackage{graphicx}
\usepackage{textcomp}
\usepackage{xcolor}

\usepackage{multirow}
\usepackage{url}
\usepackage{array}
\usepackage{natbib}

\def\BibTeX{{\rm B\kern-.05em{\sc i\kern-.025em b}\kern-.08em
    T\kern-.1667em\lower.7ex\hbox{E}\kern-.125emX}}

\begin{document}

\title{CreditARF: A Framework for Corporate Credit Rating with Annual Report and Financial Feature Integration}

\author{
\IEEEauthorblockN{Yumeng Shi\textsuperscript{1}, Zhongliang Yang*\textsuperscript{1}, DiYang Lu\textsuperscript{1}, Yisi Wang\textsuperscript{2}, Yiting Zhou\textsuperscript{3}, Linna Zhou\textsuperscript{1}}
\IEEEauthorblockA{\textit{\textsuperscript{1}School of Cyberspace Security,
Beijing University of Posts and Telecommunications,
Beijing, China} \\
\textit{\textsuperscript{2}Guotai Junan Securities,
Shanghai, China} \\
\textit{\textsuperscript{3}Hebei University of Technology,
Langfang, China} \\
yangzl@bupt.edu.cn}
}

\maketitle

\begin{abstract}
Corporate credit rating serves as a crucial intermediary service in the market economy, playing a key role in maintaining economic order. Existing credit rating models rely on financial metrics and deep learning. However, they often overlook insights from non-financial data, such as corporate annual reports. To address this, this paper introduces a corporate credit rating framework that integrates financial data with features extracted from annual reports using FinBERT, aiming to fully leverage the potential value of unstructured text data. In addition, we have developed a large-scale dataset, the Comprehensive Corporate Rating Dataset (CCRD), which combines both traditional financial data and textual data from annual reports. The experimental results show that the proposed method improves the accuracy of the rating predictions by 8–12\%, significantly improving the effectiveness and reliability of corporate credit ratings.

\end{abstract}

\section{Introduction}

Corporate credit rating, as a vital intermediary service in a market economy \cite{hilscher2017credit}, plays an indispensable role in maintaining economic stability. Corporate credit rating assesses a company’s ability to meet financial obligations, playing a crucial role in risk management. The primary objective of this evaluation is to estimate the default risk associated with the enterprise as a debtor. In addition to assisting financial professionals in mitigating potential risks, credit ratings provide investors and business partners with objective and unbiased credit information, thereby reducing operational strain on businesses.

Following the 2008 financial crisis, corporate defaults and business failures led to significant losses for both investors and financial institutions. This highlighted the importance of credit ratings in risk control. Early credit ratings relied on statistical models. With the development of machine learning technologies, methods such as Support Vector Machines (SVM) \cite{huang2004credit}, Decision Trees \cite{yeh2012hybrid}, and Ensemble Learning \cite{abellan2017comparative} have gradually been applied to this field. In recent years, neural network models, particularly Convolutional Neural Networks (CNN) \cite{feng2020every}, Recurrent Neural Networks (RNN) \cite{chen2020novel}, Graph Neural Networks (GNN) \cite{feng2022every}, and Transformer-based architectures \cite{yue2022multi}, have been widely applied to corporate credit rating problems. 

Traditionally, credit ratings have been based on financial metrics, such as liquidity ratios, profitability indicators, and debt-to-equity ratios. However, there is increasing recognition of the value of non-financial data, like corporate annual reports, industry analysis, and public sentiment. These non-financial features complement financial data and offer critical insights into a company’s operations. As a result, non-financial data, including news articles \cite{tsai2016impact}, industry reports \cite{8592626}, and credit rating action reports \cite{zhang2023investment}, have been increasingly incorporated into credit rating models, creating a more dynamic and comprehensive evaluation framework.

Among non-financial data sources, corporate annual reports have become essential for credit analysis and investment decisions \cite{balakrishnan2010predictive}. According to U.S. SEC regulations, publicly listed companies must file detailed annual reports (e.g., 10-K reports) that include financial statements, management analysis, business strategies, and potential risks. These reports address the limitations of relying solely on financial data, offering a holistic view of the challenges of a company.

However, these non-financial features are often unstructured and multi-source, making them difficult for traditional machine learning and neural network models to extract and analyze. This limits the effectiveness of these models in accurately performing corporate rating tasks.

Despite the advancements in large language models (LLMs), to the best of our knowledge, these technologies have not yet been fully applied to corporate credit ratings. In response, we propose a novel framework that integrates financial data with features from corporate annual reports. Using the power of LLMs, this framework improves the extraction and analysis of critical information, improving the accuracy and comprehensiveness of corporate credit ratings.

The primary contributions of this study are as follows:
\begin{itemize}
    \item \textbf {We developed a framework for the prediction of corporate credit rating that integrates financial data and corporate annual report data}: In this framework, we use fundamental neural networks to process traditional financial data while leveraging large language models (LLMs) to perform in-depth feature extraction from the unstructured text of annual reports. 
    \item \textbf {We constructed a large-scale corporate rating dataset that integrates traditional financial data with data from corporate annual reports}: To validate the effectiveness of the proposed framework, we developed a new dataset comprising 2,307 samples. This dataset combines 19 financial attributes with corporate annual reports, creating a comprehensive and multidimensional database.
    \item \textbf {We evaluated the generalization ability of the framework}: The results of our tests demonstrate that the incorporation of annual report features as additional data into existing corporate credit rating models significantly improves rating accuracy.
\end{itemize}

\begin{figure*}[htbp]
  \centering
  \includegraphics[width=\textwidth]{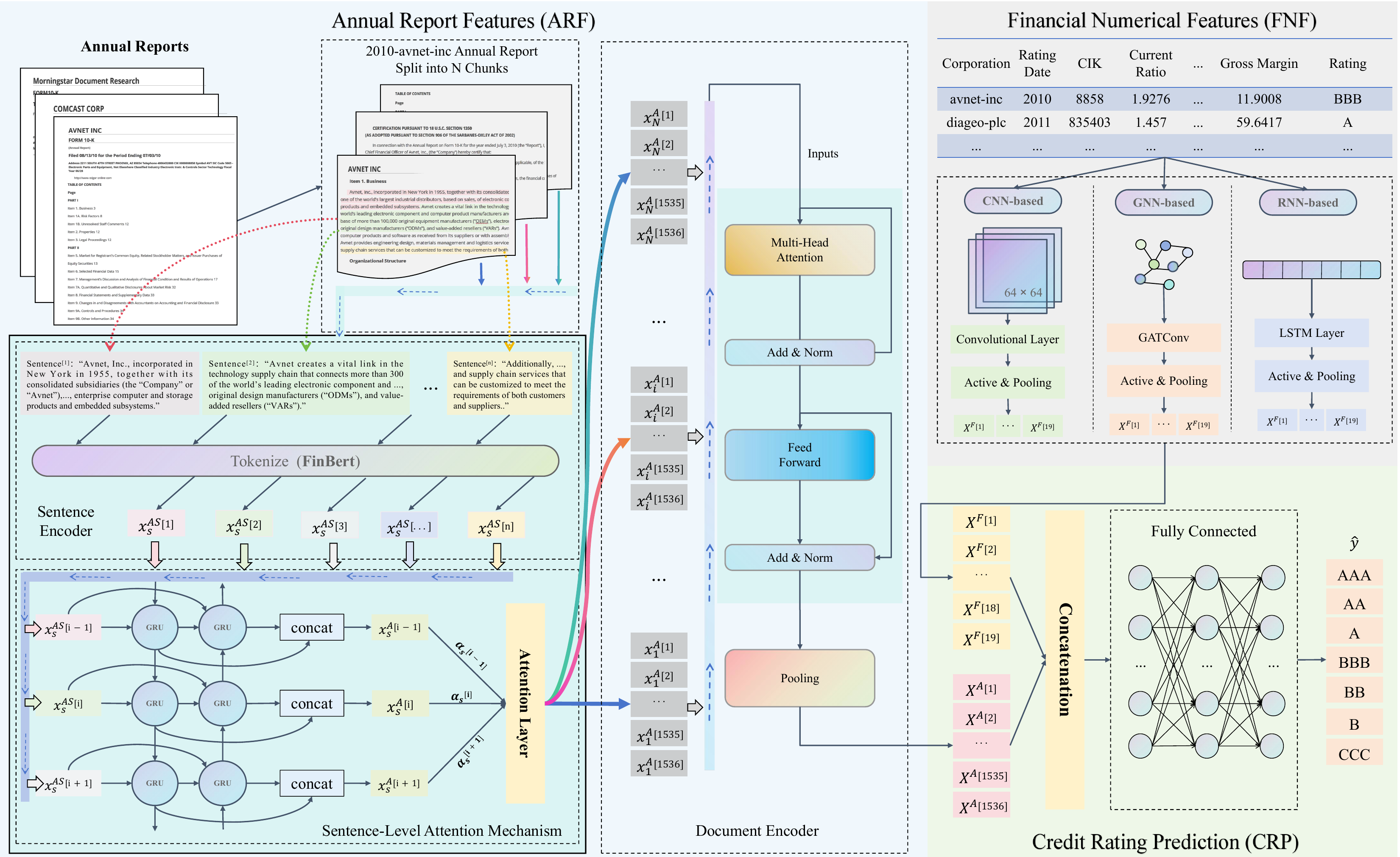} 
  \caption{Our framework consists of three main modules: Financial Feature Extraction (FNF), Annual Report Feature Extraction (ARF), and Credit Rating Prediction (CRP). The FNF module leverages traditional deep learning models to model financial data and extract key financial features. The ARF module combines FinBERT with an attention mechanism to extract deep features from the annual report text. The CRP module integrates the financial and annual report features, which are then input into a fully connected layer for classification, ultimately generating the corporate credit rating.}
  \label{fig:image-1}
\end{figure*}

\section{Related Work}

Studies on corporate credit rating models have proliferated in recent years. In 2020, Golbayani et al. \cite{golbayani2020application} pioneered the application of CNN to corporate credit rating problems. Later that year, Feng et al. \cite{feng2020every} introduced the CCR-CNN model, which uses a two-dimensional matrix to represent corporate financial information. In 2021, Feng et al. \cite{feng2021adversarial} developed the ASSL4CCR model, incorporating encoding modules and adversarial learning to improve model robustness and performance. In 2022, Feng et al. \cite{feng2022contrastive} proposed the CP4CCR model based on pre-training and self-supervised learning. This model employs a contrastive pre-training method, significantly improving the accuracy of corporate credit ratings. In the same year, Feng et al. \cite{feng2022every} applied GNN to corporate credit rating, exploring the potential of graph-structured data in credit evaluation. In 2023, Tavakoli et al. \cite{tavakoli2023multi} introduced the META model, which is based on a transformer-based autoencoder and multi-task prediction framework, further expanding the application of deep learning technologies in corporate credit rating. In 2024, Shi et al. \cite{shi2024sparsegraphsage} proposed a new method named SparseGraphSage based on GNN. This method introduces randomness into graph construction and combines diffusion and sparsity techniques, significantly improving the performance of the GraphSage model in handling complex graph data.

The research mentioned above primarily highlights the use of quantitative data in corporate ratings. It is worth noting that scholars have increasingly recognized the significance of non-financial data in corporate rating and have begun to incorporate such data into their analytical frameworks alongside financial data. In 2018, Hui et al. \cite{8592626} utilized the Sentiment Latent Dirichlet Allocation (ELDA) model to predict corporate credit risk based on public sentiment information from social media, treating news articles and social media content as non-financial data for analysis. In 2020, Choi et al. \cite{choi2020predicting} applied three methods—Bag of Words (BOW), Word2Vec, and Doc2Vec—to vectorize corporate annual reports with the aim of enhancing the predictive performance of machine learning models. In 2023, Zhang et al. \cite{zhang2023investment} built decision trees using CatBoost and LightGBM with a leaf-wise growth technique, combining both financial and non-financial data to assess the investment value of listed companies. In 2024, Chen et al. \cite{chen2024social} employed the KNN model to analyze public opinion on social media for predicting corporate credit ratings.

Although recent studies have attempted to use non-financial data from corporate annual reports for credit rating, most of these studies rely on machine learning techniques to analyze and extract information entropy from the reports. These methods have limited model capabilities and often struggle to effectively extract sufficiently rich and useful information. 

To address it, our study introduces large language models (LLMs) for deep feature extraction from unstructured text within corporate annual reports. By adopting this approach, we aim to improve the understanding of potential non-financial information within annual reports and efficiently integrate the annual report features (ARF) into financial features. 

\section{METHOD}

\subsection{Task Modeling}
To enhance the accuracy and reliability of corporate credit ratings, we propose a framework that integrates traditional financial data with features extracted from corporate annual reports. The methodology comprises three core modules: (1) Extraction of Financial Numerical Features, (2) Extraction of Annual Report Features, and (3) Credit Rating Prediction, as illustrated in Figure~\ref{fig:image-1}.

\textbf{Problem Definition}: Corporate credit rating is formulated as a multi-class classification task. Given a company \( X \), the objective is to predict a credit rating label \( \hat{y} \in \{\text{AAA}, \text{AA}, \text{A}, \text{BBB}, \text{BB}, \text{B}, \text{CCC}\} \), based on both its financial indicators and textual disclosures.

\textbf{Extraction of Financial Numerical Features}: For each company, we extract a set of structured financial indicators for the current fiscal year. These are represented as a feature vector \( \mathbf{X}^F = [x^f_1, x^f_2, \ldots, x^f_N] \), where \( x^f_i \) denotes the \( i \)-th indicator and \( N \) is the total number of financial features. A dedicated encoder \( M^{\text{FNF}} \) is employed to process \( \mathbf{X}^F \).

\textbf{Extraction of Annual Report Features}: Corporate annual reports contain essential financial and operational information. Our framework extracts key feature sets from these documents, including asset-liability conditions \( x^{\text{BS}} \), revenues and expenditures \( x^{\text{RE}} \), cash flows \( x^{\text{CF}} \), and management analysis and outlook \( x^{\text{MD}} \). These are aggregated into a textual feature vector \( \mathbf{X}^A \), which is processed by a language model \( M^{\text{ARF}} \).

\textbf{Credit Rating Prediction}: The financial and textual features are concatenated to form a unified representation: \( \mathbf{Z} = [\mathbf{X}^F; \mathbf{X}^A] \). This combined feature vector \( \mathbf{Z} \) is then fed into the credit rating prediction model \( M^{\text{CRP}} \), which outputs the predicted rating class \( \hat{y} \).

\subsection{Financial Numerical Features (FNF)} \label{FNF}
In this study, deep learning models are employed for the extraction and analysis of financial data features. Financial data typically have high-dimensional features with complex interrelationships among them. Therefore, an important task of this study is to effectively extract meaningful features from these high-dimensional data.

Previous studies have attempted to model financial data using various methods, such as models based on CNN \cite{feng2020every}, models based on GNN \cite{feng2022every}, and models based on RNN \cite{tavakoli2023multi}. To demonstrate the general applicability of the proposed framework, in this work we try to supplement the aforementioned financial numerical feature extraction models \( M^{\text{FNF}} \) with the features of the annual report extracted by large models, thus validating the effectiveness of the proposed framework.

For \textbf{CNN-based \( M^{\text{FNF}} \)}, we convert the company's embed vector \( \mathbf{x} \) from the three-channel embeddings \( \mathbf{x}_r \), \( \mathbf{x}_g \), and \( \mathbf{x}_b \) into a 2D matrix of sie \( S \times S \). Each value \( (i, j, k) \) in the matrix is calculated using the normalization formula:
\begin{equation}
\scalebox{0.7}{$ % Adjust the scaling factor (0.8) as needed
\begin{aligned}
P(i, j, k)&=\text{round} \left\{ \frac{L\left(i, (j - 1) \times S + k\right) - \text{Min}(L(i))}{\text{Max}(L(i)) - \text{Min}(L(i))} \times 255 \right\},
\end{aligned}
$}
\end{equation}
where \( j \) and \( k \) range from 1 to \( S \), and \( i \) ranges from 1 to 3. The function \( \text{round}(\cdot) \) denotes the rounding function, and the values are normalized to the range \([0, 255]\). 
Subsequently, a convolutional kernel \( \mathbf{G} \) with size \( k \times k \) is used for convolution, and the convolution result \( \mathbf{H} \) is computed as:
\begin{equation}
\mathbf{H}_{i, j} = \sum_{i=-\left\lvert \frac{k}{2} \right\rvert}^{\left\lvert \frac{k}{2} \right\rvert} \sum_{j=-\left\lvert \frac{k}{2} \right\rvert}^{\left\lvert \frac{k}{2} \right\rvert} \mathbf{P}_{:,i+m, j+n} \cdot \mathbf{G}_{i, j} + b,
\end{equation}
where $\mathbf{G}$ is the convolution kernel, and $b$ is the bias term. After several layers of convolution and pooling operations, the final result is flattened to generate the vector of financial features \( \mathbf{X}^F \).

For \textbf{GNN-based \( M^{\text{FNF}} \)}, we embed the company data into the feature representations \( \mathbf{x} \) and construct a graph \( G = (V, E) \), where \( V = \{v_1, v_2, \cdots, v_d\} \) represents the feature nodes \( d \) in the graph. Each node \( v_i \) corresponds to an attribute \(x_i \in \mathbb{R}\). In the Graph Attention Network (GAT), we simulate the varying importance between features. The feature update function at layer \( l \) is given by:
\begin{equation}
x^{(l)}_i = \alpha^{(l)}_{i,i}\Theta^{(l)}x^{(l-1)}_i + \sum_{j \in N(i)} \alpha^{(l)}_{i,j}\Theta^{(l)}x^{(l-1)}_j,
\end{equation}
where \( \Theta^{(l)} \in \mathbb{R}^{d^{(l)} \times d^{(l-1)}} \) and \( a \in \mathbb{R}^{2d^{(l)}} \) are the parameters of the GAT layer, \( x^{(l-1)} \in \mathbb{R}^{d \times d^{(l-1)}} \) is the input feature, and the attention coefficient \( \alpha^{(l)}_{i,j} \) is calculated as:
\begin{equation}
\scalebox{0.8}{$
\alpha^{(l)}_{i,j} = \frac{\exp \left( \text{LeakReLU} \left( a^{T^{(l)}} \left[ \Theta^{(l)} x^{(l-1)}_i \| \Theta^{(l)} x^{(l-1)}_j \right] \right) \right)}{\sum_{k \in N(i) \cup \{i\}} \exp \left( \text{LeakReLU} \left( a^{T^{(l)}} \left[ \Theta^{(l)} x^{(l-1)}_i \| \Theta^{(l)} x^{(l-1)}_k \right] \right) \right)}.
$}
\end{equation}

Through these operations, the financial feature vector \( \mathbf{X}^F \) generated by the GNN model captures the complex relationships between the features.

For \textbf{RNN-based \( M^{\text{FNF}} \)}, we employ a Long Short-Term Memory (LSTM) network to capture long-term dependencies within time series data. The financial data are first embedded in a high-dimensional feature space, generating an embedding vector \( x \in \mathbb{R}^{T \times \text{input\_size}} \), where \( T \) represents the number of time steps and \(\text{input\_size}\) refers to the dimension of the feature at each time step. Through these calculations, the LSTM is capable of generating the financial characteristic vector \( \mathbf{X}^F \).

\subsection{Annual Report Features (ARF)}\label{ARF}
In this study, we extract the corresponding annual reports from the database based on the year and company name, using deep learning models to extract information from PDF-formatted reports and generate document embeddings, which assist in corporate credit rating and financial analysis. The processing flow is mainly divided into two stages: pre-processing and feature extraction.

\subsubsection*{Preprocessing Stage}

In the preprocessing stage, we use specific tools such as pdfplumber and PyMuPDF to perform structured extraction of PDF files, converting the raw PDF data into text files (TXT format) to simplify subsequent processing. Specifically, these tools are used to extract text, tables, and image content from annual reports, ensuring that key structural information necessary for feature extraction is preserved.

\subsubsection*{Feature Extraction Stage}

FinBERT \cite{liu2021finbert} is a domain-specific variant derived from the highly regarded Bidirectional Encoder Representations from Transformers (BERT \cite{devlin2018bert}) model, meticulously optimized to address the unique challenges of natural language processing in the financial sector. Pre-trained on a vast corpus of financial texts, including regulatory filings (e.g., 10-K reports), earnings summaries, and market news, FinBERT is adept at capturing the subtle semantics, domain-specific terminologies, and complex syntactic patterns inherent in financial language. This targeted pre-training provides a significant advantage over general-purpose language models, enabling superior performance in tasks that require a deep understanding of financial documents.

In the present study, we utilize FinBERT to extract document-level embeddings from annual report texts, offering a robust and scalable approach for feature extraction in financial analysis. Using FinBERT’s pre-trained contextual representations, we generate high-dimensional feature vectors that preserve the semantic and syntactic nuances of the source text, supporting specific predictive tasks such as predicting corporate credit rating. The feature extraction process consists of three key components: Sentence Encoder, Sentence-Level Attention Mechanism, and Document Encoder.

In the \textbf{Sentence Encoder} part, the text is initially tokenized using the FinBERT tokenizer, which converts it into sequences of subwords or word IDs. To handle texts of varying lengths efficiently, we define the maximum sentence length as \(\mathbf{L_\text{a}}\), truncating any sequences exceeding this limit. The \(\text{batch\_size}\) parameter controls the maximum number of sentences processed per batch. In this study, the maximum sentence length is set to 512 tokens, with a maximum of 50 sentences per batch. During each iteration, the tokenized sentences are grouped into batches, and each batch is input into the FinBERT model. This process generates word embeddings, which are continuous vector representations capturing semantic and syntactic properties, as well as attention scores that assign importance to specific words. The word embeddings facilitate the understanding of relationships and contextual similarities between words, while the attention mechanism enables the model to selectively focus on salient aspects of the text.

In the \textbf{Sentence-Level Attention Mechanism} part, the [CLS] embeddings, representing the sentence embeddings, are passed through a bidirectional Gated Recurrent Unit (GRU) to capture inter-sentence dependencies and contextual information flow. This process produces context-aware sentence encodings, denoted as \(x^A_s\):
\begin{equation}
    x^A_s = \overrightarrow{\text{GRU}(x^{AS}_s)} 
    \oplus \overleftarrow{\text{GRU}(x^{AS}_s)}, \quad s = 1, 2, \dots, L_\text{a},
\end{equation}
where \(x^{AS}_s \in \mathbb{R}^{m}\) represents the [CLS] sentence embedding, \(x^A_s \in \mathbb{R}^{2m}\) denotes the context-aware sentence encoding, and \(m = 1024\) corresponds to the embedding size from FinBERT.

To compute attention scores, a linear transformation is applied to the context-aware sentence representation \(x^A_s\) using a weight matrix \(\mathbf{W} \in \mathbb{R}^{\alpha \times 2m}\) and a bias vector \(\mathbf{b} \in \mathbb{R}^{\alpha}\). In this study, we set \(\alpha = 128\) to match the attention dimension (\(\text{att\_dim} = 128\)). This transformation is followed by a non-linear activation function, such as the hyperbolic tangent (\(\tanh\)), to introduce non-linearity and constrain the output range. The resulting attention scores are computed as:

\begin{equation}
    u_s = \tanh(\mathbf{W} \cdot x^A_s + \mathbf{b}), \quad s = 1, 2, \dots, L_\text{a},
\end{equation}
where \(u_s \in \mathbb{R}^{\alpha}\), which ensures that the transformed representation has a lower dimensionality compared to the original context-aware sentence representation, \(x^A_s \in \mathbb{R}^{2m}\).

The resulting \(\mathbf{u_s}\) values are then used to compute attention weights \(\alpha_s\) via the softmax function:
\begin{equation}
    \alpha_s = \text{softmax}(\mathbf{u}_s^T \cdot \mathbf{U}), \quad s = 1, 2, \dots, L_\text{a},
\end{equation}
where \(\mathbf{U} \in \mathbb{R}^{2m}\) is a trainable attention weight vector and \(\alpha_s\) represents the attention weight assigned to each sentence.

In the \textbf{ Document Encoder} part, the encoder is based on the Transformer architecture, specifically designed to aggregate batch-level features into a fixed-dimensional document embedding. For an annual report, it is assumed to be divided into \(N\) batches. After being processed through the initial stages, each batch generates a paragraph-level feature, denoted as \(x^{A}_i\). These \(N\) paragraph-level features, \(x^{A}_i\) (\(i = 1, 2, \dots, N\)), are then input into the Transformer model. Using the capabilities of the Transformer architecture, the encoder captures both local information within individual batches and the dependencies across multiple batches. Once the Transformer has processed these features, a pooling operation is applied to aggregate them into a global document representation, denoted as \(\mathbf{X}^{A}\).

This comprehensive and expressive representation of features provides a solid foundation for the prediction of corporate credit rating.

\begin{table}[!b]
\caption{Annual Report Information for a Subset of U.S. Publicly Listed Companies}
\begin{center}
\renewcommand{\arraystretch}{1.2}
\begin{tabular}{>{\centering\arraybackslash}p{0.55\linewidth} c}
\hline
\noalign{\hrule height 0.1mm}
\textbf{Company Name} & \textbf{Years} \\
\hline
Avnet Inc. & 2010-2016 \\
Automatic Data Processing Inc. & 2010-2016 \\
Carpenter Technology Corp. & 2010-2016 \\
General Mills Inc. & 2010-2016 \\
Micron Technology Inc. & 2010-2016 \\
Compass Group PLC & 2010-2016 \\
Belden Inc. & 2011-2016 \\
AstraZeneca PLC & 2011-2016 \\
Target Corp. & 2011-2016 \\
Teleflex Incorporated & 2011-2016 \\
BCE Inc. & 2011-2016 \\
Autozone Inc. & 2011-2016 \\
% \hline
\multicolumn{2}{c}{\text{...}} \\
\hline
\noalign{\hrule height 0.1mm}
\end{tabular}
\label{tab:company_report_years}
\end{center}
\end{table}

\subsection{Credit Rating Prediction (CRP)}\label{CRP}
After completing the feature extraction of financial data and corporate annual reports, we apply a feature fusion strategy to integrate these two types of information. 

Let the financial feature vector be \( \mathbf{X}^F \), and the feature representation of the annual report be \( \mathbf{X}^A \). The concatenated feature vector \( \mathbf{Z} \) is given by the following equation:
\begin{equation}
\mathbf{Z} = [\mathbf{X}^F; \mathbf{X}^A],
\end{equation}
where \( [\cdot; \cdot] \) denotes the concatenation operation. 

Next, the concatenated high-dimensional feature vector \( \mathbf{Z} \) is fed into a fully connected layer for further feature learning and nonlinear transformation. 

To enhance the performance of the model, multiple fully connected layers can be stacked, each applying different nonlinear transformations. Generally, this process can be extended to a multi-layer perceptron (MLP), where each layer is represented as:
\begin{equation}
\mathbf{h}_k = \sigma(\mathbf{W}_k \mathbf{h}_{k-1} + \mathbf{b}_k), \quad k = 1, 2, \dots, L,
\end{equation}
where \( \mathbf{h}_0 = \mathbf{Z} \), and \( \mathbf{h}_k \) is the output of the \( k \)-th hidden layer, with \( L \) denoting the number of fully connected layers. The final predicted credit rating output can be represented as:
\begin{equation}
\hat{y} = \text{softmax}(\mathbf{W}_L \mathbf{h}_{L-1} + \mathbf{b}_L),
\end{equation}
where \( \hat{y} \) represents the predicted credit rating. The softmax function normalizes the output into a probability distribution, which is then used to determine the credit rating category, including AAA, AA, A, BBB, BB, B, and CCC.

\section{Experiment}

\subsection{Dataset}

Due to the current lack of rating datasets that integrate annual report data, we have developed the Comprehensive Corporate Rating Dataset (CCRD), a large-scale dataset that combines traditional financial data with annual report data. We obtained a dataset containing only financial data from Kaggle \footnote{\url{https://www.kaggle.com/datasets/kirtandelwadia/corporate-credit-rating-with-financial-ratios}} and then collected corresponding annual report data based on company names and years from annual report websites \footnote{\url{https://www.annualreports.com/}}. This process resulted in a new dataset containing 2,307 samples, which integrates 20 financial attributes and annual report documents. 

\begin{table}[!ht]
\caption{This table presents the primary financial features included in the dataset, covering credit ratings and financial ratio data for 5,408 U.S. publicly listed companies from 2010 to 2016.}
\begin{center}
\renewcommand{\arraystretch}{1.2}
\begin{tabular}{>{\centering\arraybackslash} p{0.6\linewidth} c }
\hline
\noalign{\hrule height 0.1mm}
\textbf{Primary Factors} & \textbf{Factor Type} \\
\hline
Rating Agency & Non-Financial \\
Corporation & Non-Financial \\
Rating & Non-Financial \\
Rating Date & Non-Financial \\
Current Ratio & Financial \\
Long-term Debt / Capital & Financial \\
Debt/Equity Ratio & Financial \\
Gross Margin & Financial \\
Operating Margin & Financial \\
EBIT Margin & Financial \\
EBITDA Margin & Financial \\
Pre-Tax Profit Margin & Financial \\
Net Profit Margin & Financial \\
Asset Turnover & Financial \\
ROE - Return On Equity & Financial \\
Return On Tangible Equity & Financial \\
ROA - Return On Assets & Financial \\
ROI - Return On Investment & Financial \\
Operating Cash Flow Per Share & Financial \\
Free Cash Flow Per Share & Financial \\
% \hline
\multicolumn{2}{c}{\text{...}} \\
\hline
\noalign{\hrule height 0.1mm}
\end{tabular}
\label{tab:financial_features}
\end{center}
\end{table}

\begin{table*}[htpb]
\caption{The table presents the experimental results of multiple models, including traditional baseline models, and their performance changes after incorporating the annual report feature (ARF).}
\begin{center}
\renewcommand{\arraystretch}{1.2}
\begin{tabular}{>{\centering\arraybackslash} p{0.25\linewidth}|>{\centering\arraybackslash}p{0.05\linewidth}|>{\centering\arraybackslash}p{0.08\linewidth}| >{\centering\arraybackslash}p{0.05\linewidth} >{\centering\arraybackslash}p{0.05\linewidth} >{\centering\arraybackslash}p{0.05\linewidth} >{\centering\arraybackslash}p{0.05\linewidth} >{\centering\arraybackslash}p{0.05\linewidth} >{\centering\arraybackslash}p{0.05\linewidth} >{\centering\arraybackslash}p{0.05\linewidth}}
\hline
\noalign{\hrule height 0.1mm}
\textbf{Models} & & \textbf{All} & \textbf{AAA} & \textbf{AA} & \textbf{A} & \textbf{BBB} & \textbf{BB} & \textbf{B} & \textbf{CCC}\\
\hline
\multirow{3}{*}{LR \cite{feng2020every}} & \textbf{Rec} & 0.248& 0.843& 0.049& 0.000& 0.022& 0.006& 0.704& 0.056\\
 & \textbf{Acc} & 0.248& 0.154& 0.155& 0.145& 0.139& 0.129& 0.141& 0.136\\
 & \textbf{F1} & 0.147& 0.460& 0.085& 0.000 & 0.037& 0.011& 0.329& 0.072\\
\hline
\multirow{3}{*}{SVM \cite{feng2020every}} & \textbf{Rec} & 0.540& 0.441& 0.502& 0.469& 0.527& 0.474& 0.414& 0.978\\
 & \textbf{Acc} & 0.540& 0.154& 0.155& 0.145& 0.139& 0.129& 0.141& 0.136\\
 & \textbf{F1} & 0.563& 0.575& 0.628& 0.563& 0.591& 0.589& 0.546& 0.441\\
\hline
\hline
\multirow{3}{*}{(Feng et al., 2020) \cite{feng2020every}} & \textbf{Rec} & 0.727 & 0.873 & 0.811 & 0.546 & 0.625 & 0.719 & 0.621 & 0.904 \\
 & \textbf{Acc} & 0.727 & 0.932& 0.662& 0.537& 0.703& 0.732& 0.719& 0.805\\
 & \textbf{F1} & 0.725 & 0.902 & 0.729 & 0.541 & 0.662 & 0.726 & 0.667 & 0.852 \\
\hline
\multirow{3}{*}{(Feng et al., 2020) \cite{feng2020every} + ARF} & \textbf{Rec} & \textbf{0.817}& \textbf{0.992}& \textbf{0.818}& \textbf{0.604}& \textbf{0.780}& \textbf{0.793}& \textbf{0.849}& \textbf{0.947}\\
 & \textbf{Acc} & \textbf{0.817}& \textbf{0.937}& \textbf{0.914}& \textbf{0.643}& \textbf{0.739}& \textbf{0.780}& \textbf{0.830}& \textbf{0.941}\\
 & \textbf{F1} & \textbf{0.816}& \textbf{0.964}& \textbf{0.863}& \textbf{0.623}& \textbf{0.759}& \textbf{0.787}& \textbf{0.840}& \textbf{0.944}\\
\hline
\hline
\multirow{3}{*}{(Feng et al., 2022) \cite{feng2022every}} & \textbf{Rec} & 0.698& 0.975& 0.639& 0.427& 0.652& 0.737& 0.532& 0.922\\
 & \textbf{Acc} & 0.698& 0.799& 0.708& 0.554& 0.591& 0.670& 0.756& 0.761\\
 & \textbf{F1} & 0.688& 0.879& 0.672& 0.482& 0.620& 0.702& 0.625& 0.834\\
\hline
\multirow{3}{*}{(Feng et al., 2022) \cite{feng2022every} + ARF} & \textbf{Rec} & \textbf{0.817}& \textbf{0.985}& \textbf{0.829}& \textbf{0.604}& \textbf{0.679}& \textbf{0.807}& \textbf{0.823}& \textbf{0.983}\\
 & \textbf{Acc} & \textbf{0.817}& \textbf{0.901}& \textbf{0.806}& \textbf{0.659}& \textbf{0.735}& \textbf{0.879}& \textbf{0.805}& \textbf{0.908}\\
 & \textbf{F1} & \textbf{0.814}& \textbf{0.941}& \textbf{0.817}& \textbf{0.630}& \textbf{0.706}& \textbf{0.841}& \textbf{0.814}& \textbf{0.944}\\
\hline
\hline
\multirow{3}{*}{(Tavakoli et al., 2023) \cite{tavakoli2023multi}} & \textbf{Rec} & 0.730& 0.927& 0.751& 0.459& 0.792& 0.766& 0.538& 0.950\\
 & \textbf{Acc} & 0.730& 0.841& 0.770& 0.601& 0.681& 0.724& 0.719& 0.784\\
 & \textbf{F1} & 0..721& 0.882& 0.761& 0.521& 0.732& 0.744& 0.615& 0.859\\
\hline
\multirow{3}{*}{(Tavakoli et al., 2023) \cite{tavakoli2023multi} + ARF} & \textbf{Rec} & \textbf{0.829}& \textbf{0.993}& \textbf{0.870}& \textbf{0.565}& \textbf{0.792}& \textbf{0.865}& \textbf{0.817}& \textbf{0.989}\\
 & \textbf{Acc} & \textbf{0.829}& \textbf{0.894}& \textbf{0.858}& \textbf{0.688}& \textbf{0.736}& \textbf{0.897}& \textbf{0.864}& \textbf{0.904}\\
 & \textbf{F1} & \textbf{0.825}& \textbf{0.941}& \textbf{0.864}& \textbf{0.621}& \textbf{0.763}& \textbf{0.881}& \textbf{0.840}& \textbf{0.944}\\
\hline
\noalign{\hrule height 0.1mm}
\end{tabular}
\end{center}
\label{tab:modelperformance}
\end{table*}

The dataset we have constructed specifically contains two main parts:
\begin{itemize}
    \item \textbf{Annual Report Data}: The second component consists of 1,329 annual reports. Each annual report covers a specific year between 2010 and 2016, with lengths ranging from a few pages to more than 150 pages. Table~\ref{tab:company_report_years} displays the information on the annual report for a selection of publicly listed US companies from the data set.
    \item \textbf{Financial Data}: This part consists of 5,408 instances, which include credit ratings provided by six major rating agencies, such as Standard \& Poor's Ratings Services, Egan-Jones Ratings Company, and Moody's Investors Service, among others. The dataset pertains to the credit ratings of publicly listed US companies from 2010 to 2016. The original rating scale includes 23 grades, such as AAA, AA, BBB, etc. Specific financial features are detailed in Table~\ref{tab:financial_features}.
\end{itemize}

To optimize the experimental analysis and address the issue of limited samples in certain rating categories, the 23 rating grades were consolidated into 7 major categories: AAA, AA, A, BBB, BB, B, and CCC. Furthermore, the data from the annual report were merged with the financial data, resulting in a unified dataset comprising 2,307 instances.

\subsection{Baseline Methods}
We compare our methods with three baseline models:
\begin{itemize}
    \item \textbf{(Feng et al., 2020) \cite{feng2020every}:} This study proposed traditional machine learning models, including LR and SVM, for credit rating and also introduced CNN to leverage complex patterns in the data.
    \item \textbf{(Feng et al., 2022) \cite{feng2022every}:} This study investigated the application of GNN to improve the effectiveness of credit rating predictions using advanced deep learning techniques.
    \item \textbf{(Tavakoli et al., 2023) \cite{tavakoli2023multi}:} This study used Long Short-Term Memory (LSTM) networks to enhance the prediction of corporate credit ratings by integrating both structured and unstructured data.
\end{itemize}

\subsection{Hyperparameter Setup}
In this study, the dataset was split into a 75\% training set and a 25\% test set. To address the issue of class imbalance and ensure a balanced distribution of samples across different categories, we applied the synthetic minority over-sampling technique (SMOTE) for data augmentation in all experiments. During model training, the Adam optimizer was used, with the learning rate initialized at 0.001 and a weight decay coefficient of 0.00001 to prevent overfitting. Additionally, a learning rate scheduler, ReduceLROnPlateau, was used to reduce the learning rate by a factor of 0.5 when the validation loss plateaued for 3 consecutive epochs. The minimum learning rate was set to 1e-6, ensuring stable training throughout the process. 

\subsection{Experimental Results}
The experimental results demonstrate a significant improvement in model performance for corporate credit rating prediction tasks by incorporating annual report data extracted using large language models (LLMs). Specifically, the key findings are as follows:
\begin{itemize}
    \item After adding annual report features (ARF), the accuracy improved by approximately 8\% to 12\% across all models.
    \item Among the seven rating categories, the BB category showed the largest precision improvement, exceeding 20\%.
    \item Among the three baseline models, the GNN-based models benefited the most from the integration of ARF features, with an improvement close to 12\%.
\end{itemize}

Table~\ref{tab:modelperformance} summarizes the experimental results, comparing various models, including traditional machine learning models (such as LR and SVM) and deep learning models (such as CNN, GNN, and RNN). It also presents the performance changes after incorporating the annual report data into these baseline models. The evaluation metrics used include recall, accuracy, and F1 score, with model performance analyzed for both general data and specific credit rating categories (AAA, AA, A, BBB, BB, B, and CCC).

Firstly, compared to traditional machine learning models, neural network models (Feng et al., 2020) \cite{feng2020every}, Feng et al. (2022) \cite{feng2022every}, and Tavakoli et al. (2023) \cite{tavakoli2023multi} exhibit superior processing capabilities, effectively capturing the complex structure of financial data and demonstrating better performance. This advantage stems from the neural networks' ability to handle nonlinear relationships and extract deep features from data, which is particularly crucial in financial data analysis.

Secondly, after incorporating the 1536-dimensional annual report features generated by large language models into the existing models, we observed a significant improvement in model performance. Specifically, after adding the Corporate Annual Report Features (ARF), the overall rating accuracy of the method proposed by Feng et al. (2020) increased by nearly 8\%, Feng et al. (2022) improved by nearly 12\%, and Tavakoli et al. (2023) showed a 10\% improvement. This indicates that the annual report vectors generated by large models substantially enhance classification performance.

Further analysis of the experimental results reveals that the incorporation of ARF led to significant improvements in model performance across various rating categories for the models proposed by Feng et al. (2020), Feng et al. (2022), and Tavakoli et al. (2023). In the model by Feng et al. (2020), the precision for the AA category increased by over 25\%, with the lowest increase in the AAA category reaching 0.5\%. The average improvement across the seven categories was 9.91\%, with the highest precision for CCC at 94.1\%. In the model by Feng et al. (2022), the precision for the BB category increased by over 20\%, with the lowest increase in the B category reaching 4.9\%. The average improvement across the seven categories was 12.2\%, with the highest precision for CCC at 90.8\%. In the model by Tavakoli et al. (2023), the precision for the BB category increased by more than 17\%, with the lowest increase in the AAA category reaching 5.3\%. The average improvement across the seven categories was 10.3\%, with the highest precision for CCC at 90.4\%.

\begin{figure*}[htbp]
    \centering
    \resizebox{1.00\textwidth}{!}{ % 缩放比例：100%
    \begin{minipage}{\textwidth}
        \centering
        \begin{minipage}{0.32\textwidth}
            \centering
            \includegraphics[width=\linewidth]{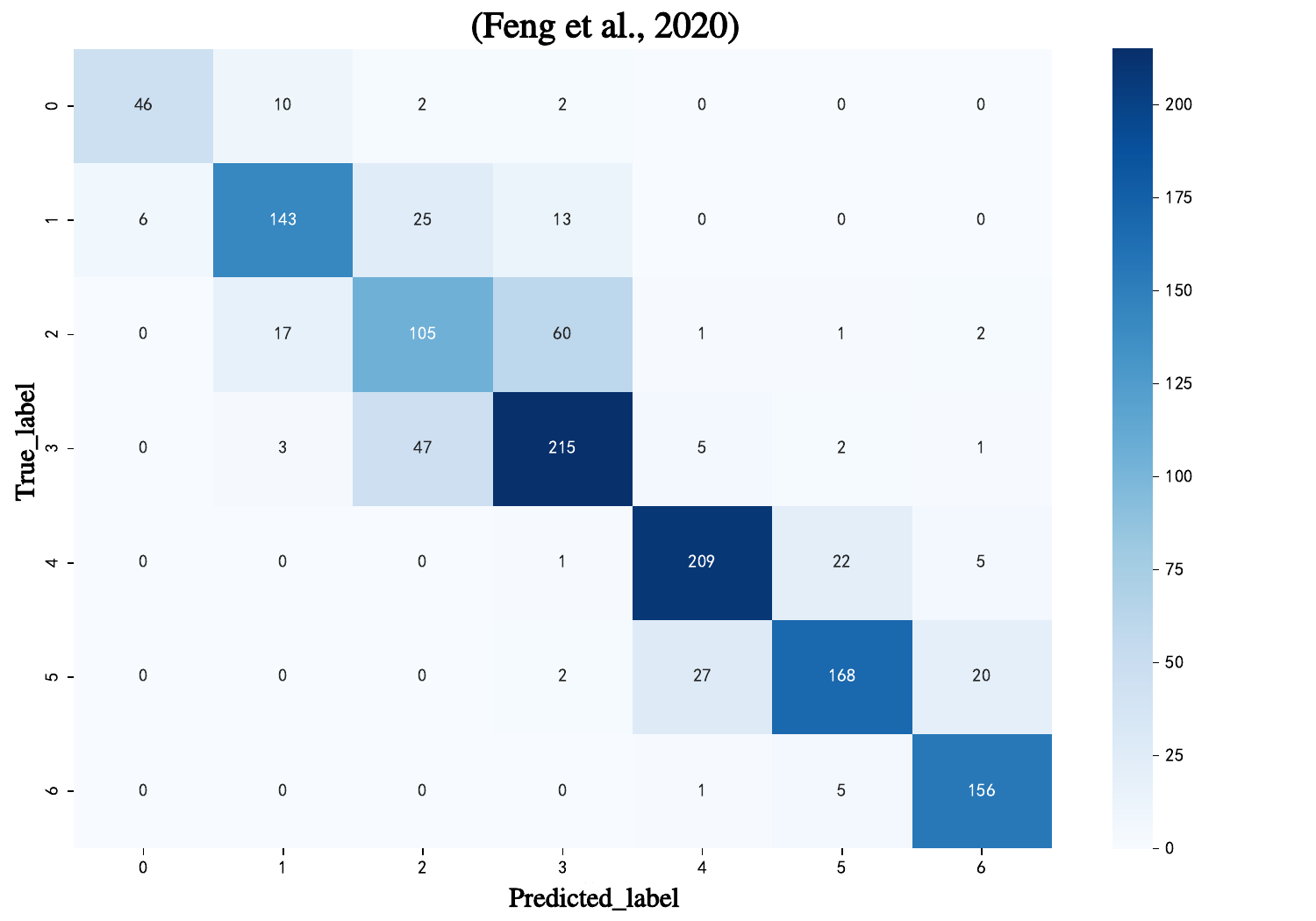}
            \label{fig:confusion2020}
        \end{minipage}%
        \hfill
        \begin{minipage}{0.32\textwidth}
            \centering
            \includegraphics[width=\linewidth]{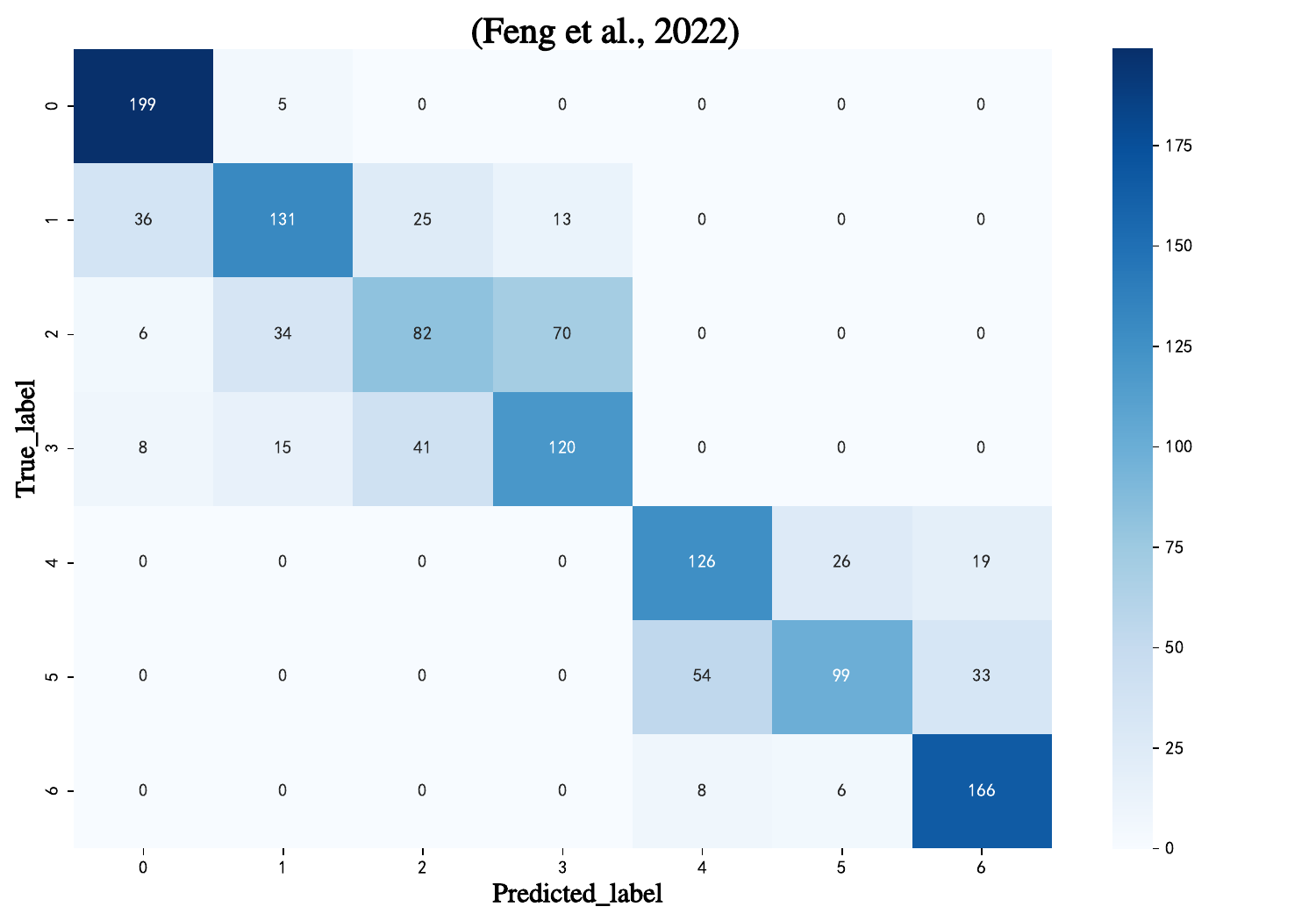}
            \label{fig:confusion2022}
        \end{minipage}%
        \hfill
        \begin{minipage}{0.32\textwidth}
            \centering
            \includegraphics[width=\linewidth]{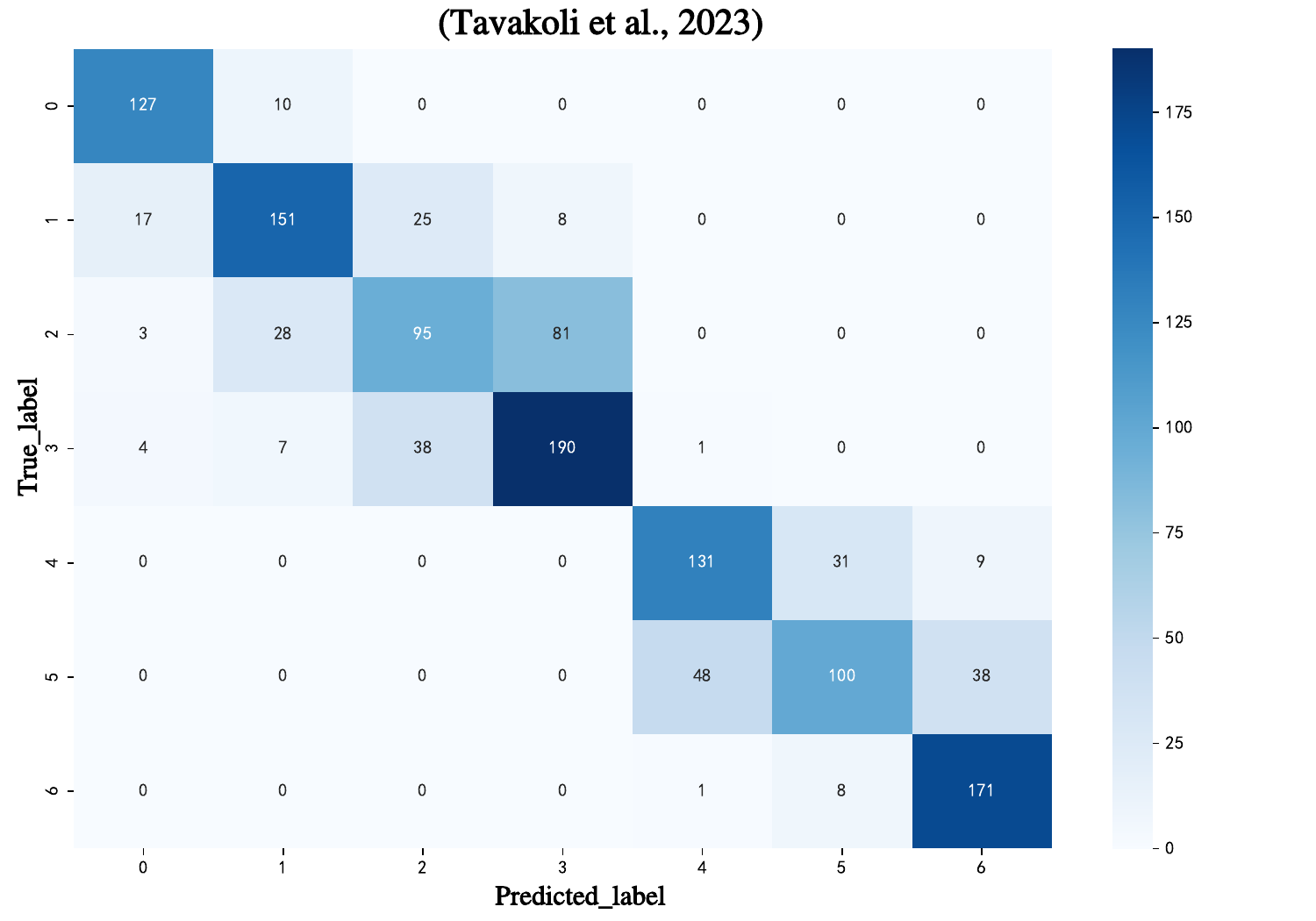}
            \label{fig:confusion2023}
        \end{minipage}
        \par\vspace{1em}
        \begin{minipage}{0.32\textwidth}
            \centering
            \includegraphics[width=\linewidth]{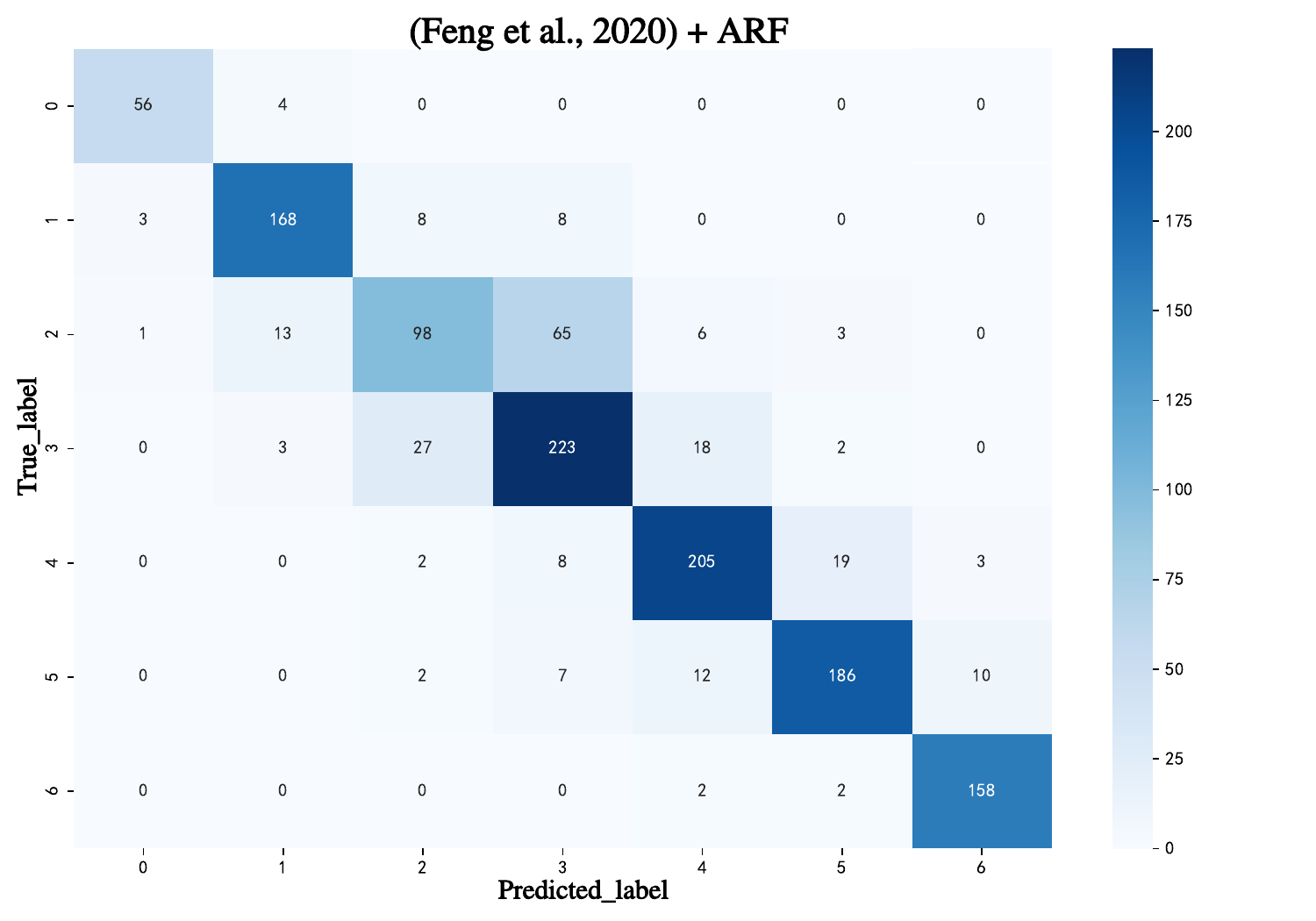}
            \label{fig:confusion2020}
        \end{minipage}%
        \hfill
        \begin{minipage}{0.32\textwidth}
            \centering
            \includegraphics[width=\linewidth]{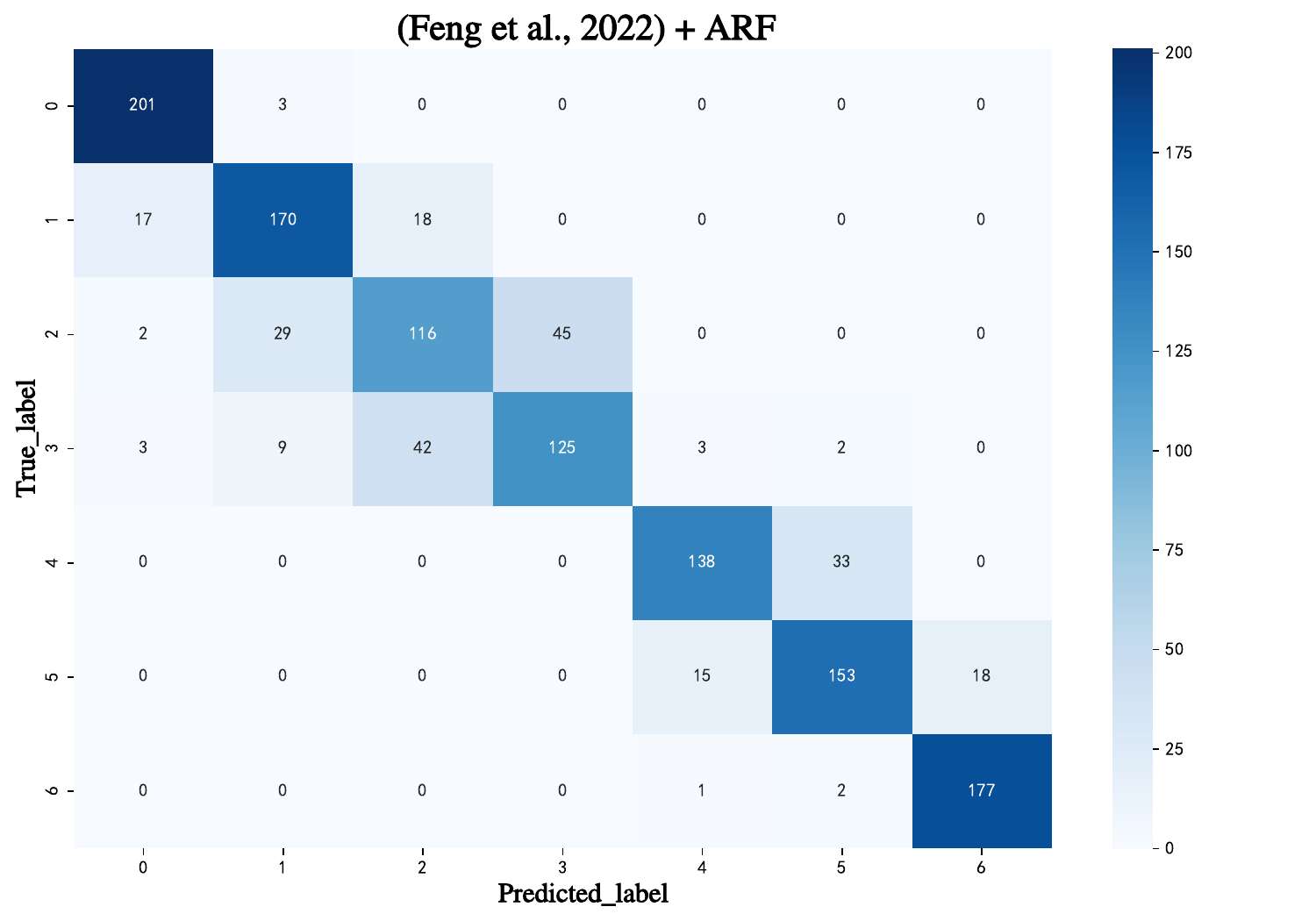}
            \label{fig:confusion2022}
        \end{minipage}%
        \hfill
        \begin{minipage}{0.32\textwidth}
            \centering
            \includegraphics[width=\linewidth]{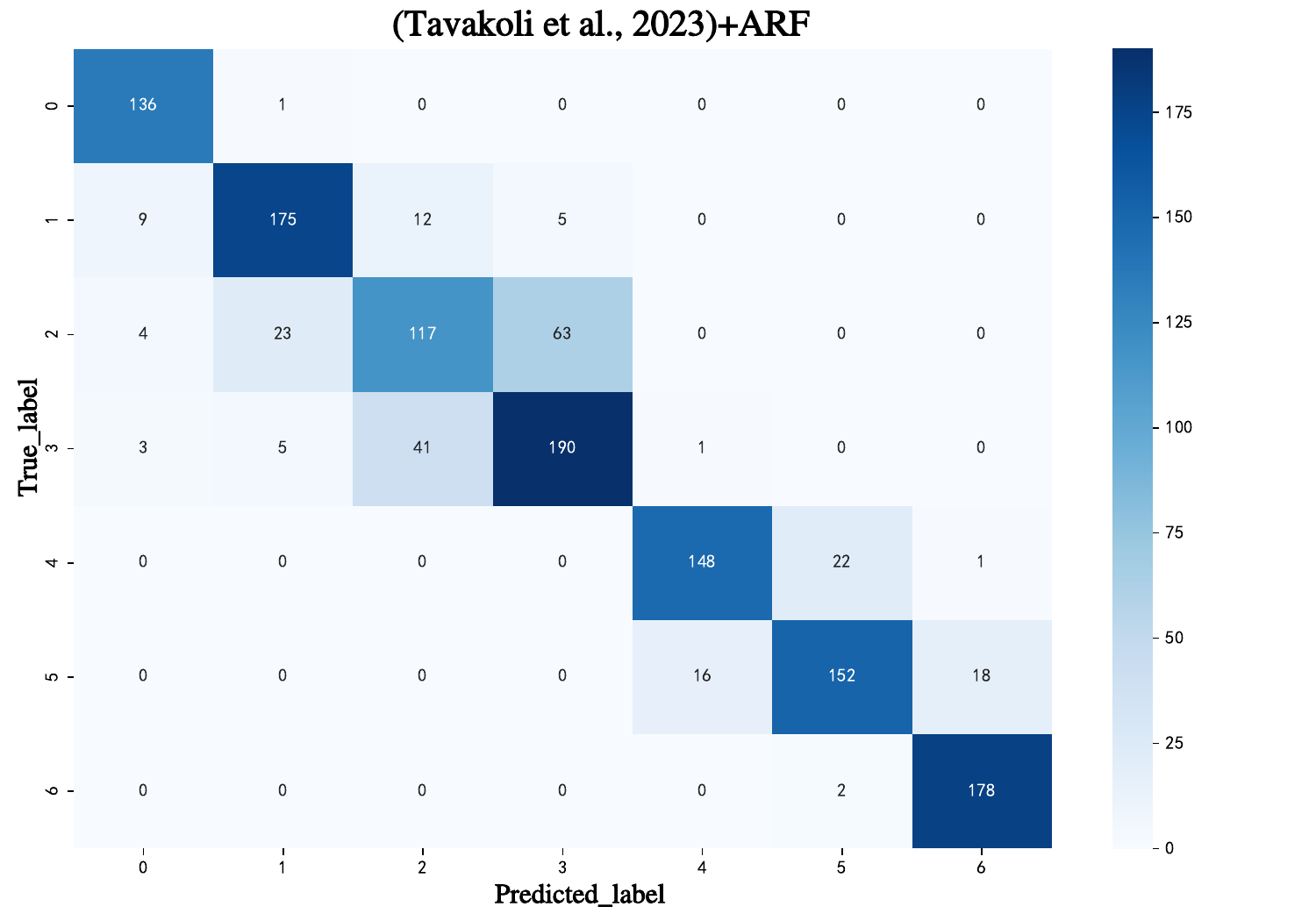}
            \label{fig:confusion2023}
        \end{minipage}
    \end{minipage}
    }
    \caption{Overall comparison of the confusion matrices for the models from \cite{feng2020every}, \cite{feng2022every}, and \cite{tavakoli2023multi}, illustrating the model performance on the test set before and after incorporating corporate annual report features (ARF). The category labels 0 to 6 in the confusion matrices correspond to AAA, AA, A, BBB, BB, B, and CCC, respectively.}
    \label{fig:confusion-matrices}
\end{figure*}

These results demonstrate that the incorporation of ARF significantly improves the model's predictive capability across various rating levels, highlighting the effectiveness of utilizing annual report features to enhance predictive performance. Moreover, the comparison of confusion matrices in the experiment (as shown in Figure~\ref{fig:confusion-matrices}) clearly illustrates that the predictive ability of the model was significantly improved after integrating the features of the corporate annual report.

\section{Conclusion}
This study proposes a novel framework for the prediction of corporate credit rating that integrates financial indicators with features extracted from annual reports. By leveraging large language models (LLMs) to extract textual information from annual reports and combining it with traditional financial data, experimental results demonstrate that the proposed method yields significant performance improvements across various neural network models. The accuracy of the model improved by 8–12\%, with the GNN-based model showing the most pronounced gains.

\section*{Acknowledgment}
This work was supported in part by the National Key Research and Development Program of China under Grant 2023YFC3305401, and in part by the National Natural Science Foundation of China (Nos.62302059 and 62172053).

% \end{thebibliography}
\bibliographystyle{unsrt}
\bibliography{IEEEabrv,custom}
\end{document}